\documentclass[12pt]{article}
\usepackage[paper=a4paper,dvips,top=2.5cm,left=2.5cm,right=2.5cm,
    foot=2cm,bottom=2.5cm]{geometry}
\usepackage{hyperref}
\hypersetup{
    colorlinks=true,
    linkcolor=blue,
    filecolor=blue,      
    urlcolor=blue,
    citecolor=blue,
    pdfpagemode=FullScreen,
    }
\usepackage{amsmath,amssymb}

\DeclareMathOperator*{\argmin}{arg\,min}
\usepackage{graphicx}
\usepackage{appendix}
\usepackage{blindtext}
\usepackage{subfiles}
\usepackage{cite}
\usepackage{multirow}
\usepackage{subcaption}
\usepackage{booktabs}
\usepackage{tabularx}
\usepackage{siunitx}
\usepackage{caption}
\usepackage[usenames,dvipsnames]{color}
\usepackage{authblk}
\usepackage{makecell}
\usepackage{array}
\usepackage{xcolor}

\title{Rapid Estimation of Left Ventricular Contractility with a Physics-Informed Neural Network Inverse Modeling Approach}

\author[1]{Ehsan Naghavi}
\author[1]{Haifeng Wang}
\author[2]{Lei Fan}
\author[3]{Jenny S. Choy}
\author[3]{Ghassan Kassab}
\author[1]{Seungik Baek}
\author[1]{Lik-Chuan Lee \thanks{Corresponding author: lclee@egr.msu.edu}}
\affil[1]{Department of Mechanical Engineering, Michigan State University, East Lansing, MI}
\affil[2]{Joint Department of Biomedical Engineering, Marquette University and Medical College of Wisconsin, Milwaukee, WI}
\affil[3]{California Medical Innovations Institute, San Diego, CA}
\date{}

\begin{document}

\maketitle

\section*{Abstract}
Physics-based computer models based on numerical solution of the governing equations generally cannot make rapid predictions, which in turn, limits their applications in the clinic. 
To address this issue, we developed a physics-informed neural network (PINN) model that encodes the physics of a closed-loop blood circulation system embedding a left ventricle (LV).
The PINN model is trained to satisfy a system of ordinary differential equations (ODEs) associated with a lumped parameter description of the circulatory system.
The model predictions have a maximum error of less than $5\%$ when compared to those obtained by solving the ODEs numerically.
An inverse modeling approach using the PINN model is also developed to rapidly estimate model parameters (in $\sim$ 3 mins) from single-beat LV pressure and volume waveforms.
Using synthetic LV pressure and volume waveforms generated by the PINN model with different model parameter values, we show that the inverse modeling approach can recover the corresponding ground truth values, which suggests that the model parameters are unique. 
The PINN inverse modeling approach is then applied to estimate LV contractility indexed by the end-systolic elastance $E_{es}$ using waveforms acquired from 11 swine models, including waveforms acquired before and after administration of dobutamine (an inotropic agent) in 3 animals. 
The estimated $E_{es}$ is about 58\% to 284\% higher for the data associated with dobutamine compared to those without, which implies that this approach can be used to estimate LV contractility using single-beat measurements. 
The PINN inverse modeling can potentially be used in the clinic to simultaneously estimate LV contractility and other physiological parameters from single-beat measurements. 

\hfill

\noindent \textit{Keywords:} Cardiac contractility, Patient-specific Modeling, Physics-informed Neural Network, Lumped Parameter Model, Parameter Estimation, Sensitivity Analysis

\section{Introduction}

Computational and/or mathematical models are increasingly used as patient-specific tools to understand cardiovascular diseases and treatments \cite{Figueroa2009, Gray2018, Niederer2019, Schwarz2023} as well as to quantify changes in heart function (e.g., cardiac contractility)  \cite{Charoenpanichkit2010}. 
These models are usually developed based on numerical methods (e.g., finite element method) that are used to solve the partial differential equations (PDEs) and/or ordinary differential equations (ODEs) governing the key driving physics. Such models are often referred to as ``physics-based'' models \cite{MIGLIAVACCA20061010, Kim2010PatientSpecificMO, xiao2014systematic, Boileau2018, bertaglia2020computational, Shavik2020, Muller2021, Zambrano2021, grande2022hybrid, PATEL2023, SCHWARZ2023116312}. 
Personalizing a physics-based model for each patient, however, requires estimating its parameters through inverse modeling, a process that often necessitates multiple runs of the numerical model. This process is time-consuming and highly computationally expensive, making it impractical when dealing with a large number of patient datasets or in time-sensitive clinical situations \cite{Baek2022, LIANG2023116347}.

Machine learning (ML) models have emerged as a promising solution to issues related to the lengthy computational time associated with patient-specific modeling \cite{Arzani2022} and modeling of biological systems \cite{maher2019accelerating,Alber2019, Dabiri2019, Dabiri2020, Rasheed2020, peng2021, KONG2021102222}. While most ML models are trained purely with substantial amount of labeled data (i.e., data-driven ML model), another class of ML models that are directly trained to satisfy the governing equations of the operating physics (i.e., physics-based ML model) is increasingly being developed. Specifically, physics-informed neural networks (PINNs) where residuals of the governing equations are incorporated in the loss function during training with a small set of data or without data are increasingly developed \cite{RAISSI2019} and applied in various biomechanical studies  \cite{KISSAS2020112623, SUN2020112732, yazdani2020systems, GAO2021110079, YIN2021113603, ArzaniNearWall2021, DALTON2023116351, LIANG2023116347, Wong2023}. To the best of our knowledge, a PINN model describing hemodynamics in the cardiovascular system with varying cardiac function has not been developed.

Here, we developed a PINN modeling framework for rapid prediction of hemodynamics in the closed-loop cardiovascular system that is described by a system of ODEs. 
We use the PINN model to perform global sensitivity analysis that would otherwise be computationally prohibitive if numerical methods were used to solve the ODEs. 
We also apply the PINN framework in an inverse modeling approach to rapidly estimate physiologically-relevant parameters such as the left ventricle contractility from animal-specific hemodynamics waveforms.  
We demonstrate the translational potential of the PINN inverse modeling approach as a clinical tool for quantifying heart function by showing that this approach can capture an increase in left ventricle contractility with hemodynamics waveforms acquired after the administration of dobutamine.

\section{Methods}

  \subsection{Mathematical model}
  \subsubsection{Blood circulation}
\begin{figure}
        \centering
        \includegraphics[width=\textwidth]{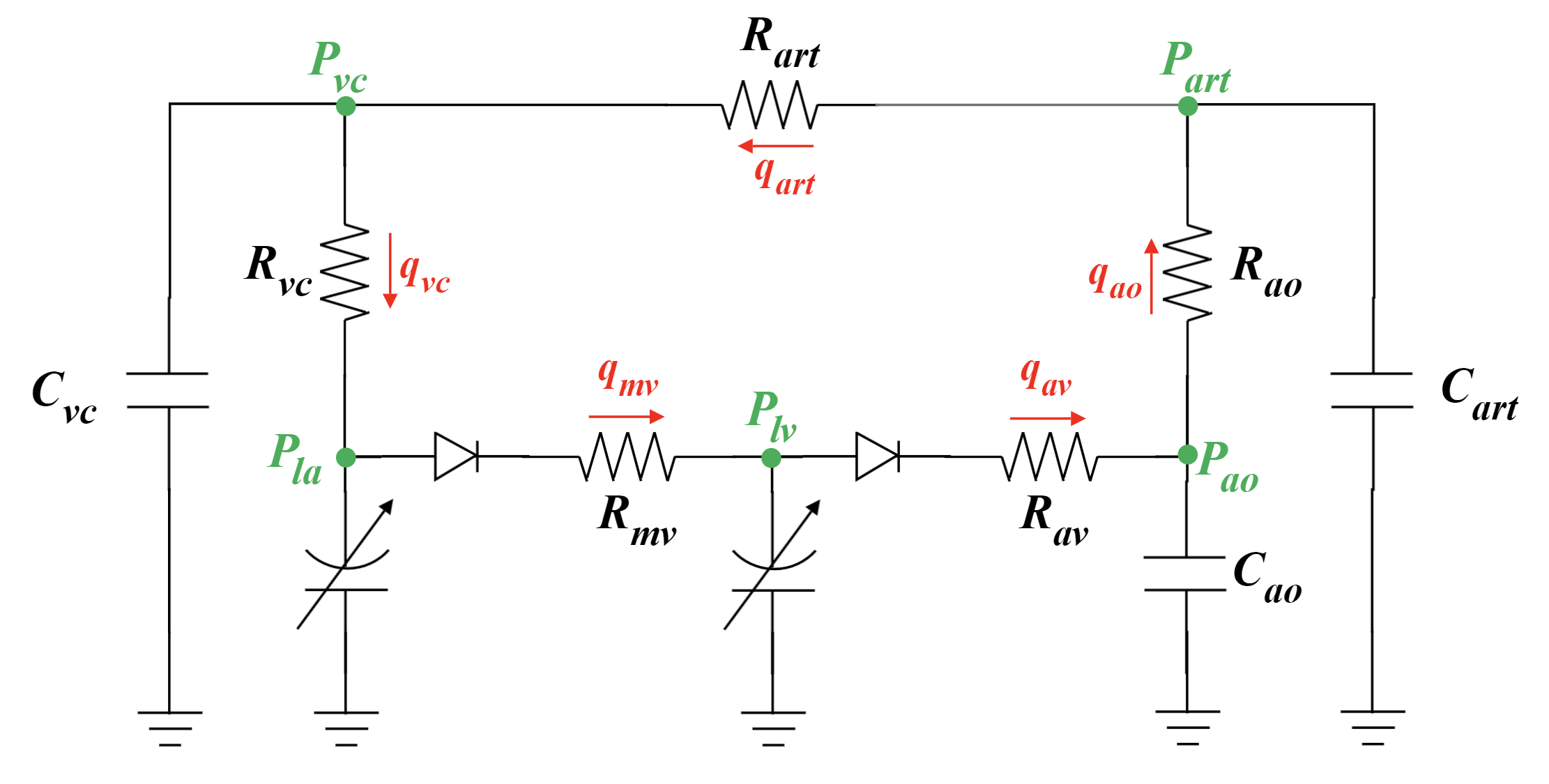}
        \caption{ The electrical equivalent diagram of the closed-loop blood circulation. lv, left ventricle; ao, aorta; art, peripheral artery; vc, vena cava; la, left atrium; av, aortic valve; mv, mitral valve.}\label{fig:Circuit}
    \end{figure}

    The closed-loop lumped parameter circulatory model is comprised of five compartments (\autoref{fig:Circuit}). The volume change of each compartment is governed by inflow and outflow rates as described by the following system of ODEs
  
    \begin{subequations} \label{eqn:ODEs}
        \begin{align}
          \frac{dV_{lv}}{dt} &= q_{mv} - q_{av}\,, \\
          \frac{dV_{ao}}{dt} &= q_{av} - q_{ao}\,,  \\
          \frac{dV_{art}}{dt} &= q_{ao} - q_{art}\,,  \\
          \frac{dV_{vc}}{dt} &= q_{art} - q_{vc}\,,  \\
          \frac{dV_{la}}{dt} &= q_{vc} - q_{mv}\,, 
        \end{align}
    \end{subequations}
    where $V_{lv}$, $V_{ao}$, $V_{art}$, $V_{vc}$, $V_{la}$ are volumes of the left ventricle (lv), aorta (ao), peripheral arteries (art), vena cava (vc), and left atrium (la), respectively. Flow rates in the aorta $q_{ao}$, peripheral arteries $q_{art}$ and vena cava $q_{vc}$, as well as across the mitral valve $q_{mv}$ and aortic valve $q_{av}$ are determined from their corresponding resistances (i.e., $R_{ao}$, $R_{art}$, $R_{vc}$, $R_{mv}$, $R_{av}$) and pressure differences between the connected storage compartments by
    \begin{subequations} \label{eqn:flowrates}
      \begin{eqnarray}
        q_{ao} &= &\dfrac{P_{ao} - P_{art}}{R_{ao}}\,, \label{eqn:flowrates_ao} \\
        q_{art} &= &\dfrac{P_{art} - P_{vc}}{R_{art}}\,, \label{eqn:flowrates_art} \\
        q_{vc} &= &\dfrac{P_{vc} - P_{la}}{R_{vc}}\,, \label{eqn:flowrates_vc} \\
        q_{av} &= &\dfrac{\max(P_{lv} - P_{ao},\, 0)}{R_{av}}\,, \label{eqn:flowrates_av} \\
        q_{mv} &= &\dfrac{\max(P_{la} - P_{lv},\, 0)}{R_{mv}}\,. \label{eqn:flowrates_mv}
      \end{eqnarray}
    \end{subequations}

    Note that Eqs. (\ref{eqn:flowrates_av}) and (\ref{eqn:flowrates_mv}) are piecewise equations  describing the flow rate by the valves' opening and closing. These equations do not exhibit smoothness in their derivatives, which can pose an issue when training the PINN model. To circumvent this issue, a smooth approximation of the maximum function given by
    \begin{align}\label{eqn:smoothmax}
      \max(x - y,\, 0) \approx \dfrac{1}{\alpha} \text{softplus}\left(\alpha(x-y)\right)\,,
    \end{align}
    is used instead, where
    \begin{align} \label{eqn:softplus}
      \text{softplus}(\alpha(x-y)) =
      \begin{cases}
        \log(1 + \exp(\alpha(x-y))); & \alpha(x-y) \le 20 \\
        \alpha(x-y); & \alpha(x-y) > 20\,,
      \end{cases}
    \end{align} 
    and $\alpha$ is a parameter that controls the smoothness of the function approximation. Larger values of $\alpha$ lead to a closer approximation to the actual maximum function. Pressure in the blood vessels is given by
    \begin{subequations} \label{eqn:pressures}
      \begin{eqnarray}
        P_{ao} &= &\dfrac{V_{ao} - V_{ao,\, r}}{C_{ao}}\,, \\
        P_{art} &= &\dfrac{V_{art} - V_{art,\, r}}{C_{art}}\,, \\
        P_{vc} &= &\dfrac{V_{vc} - V_{vc,\, r}}{C_{vc}}\,,
      \end{eqnarray}
    \end{subequations}
    where $V_{ao,\, r}$, $V_{art,\, r}$, and $V_{vc,\, r}$ are the prescribed resting volumes of the aorta, peripheral arteries, and vena cava, respectively. The corresponding prescribed compliance of the compartments are $C_{ao}$, $C_{art}$, and $C_{vc}$.
  
  \subsubsection{Heart model}
    Contraction of the left ventricle is described using a time-varying elastance model \cite{Santamore1991, witzenburg2018predicting} given as
    \begin{align}
      P_{lv}(V_{lv}, t) = e(t) P_{es}(V_{lv}) + (1 - e(t)) P_{ed}(V_{lv})\,,
    \end{align}
    where $P_{es}$ and $P_{ed}$ represent the end-systolic and end-diastolic pressure-volume relationship, respectively. These relationships are defined by
    \begin{subequations}\label{eqn:pv_rels}
      \begin{eqnarray}
        P_{es}(V_{lv}) &= &E_{es, \, lv} (V_{lv} -V_{lv, \, r})\,, \\
        P_{ed}(V_{lv}) &= &A_{lv} \left( \exp\left[B_{lv}(V_{lv} - V_{lv, \,r})\right] - 1 \right)\,,
      \end{eqnarray}
    \end{subequations}
    where $E_{es,\, lv}$ represents the end-systolic elastance of the left ventricle, $V_{lv,\, r}$ is the unloaded volume of the ventricle at end-systole, whereas $A_{lv}$ and $B_{lv}$ are coefficients that describe the exponential shape of the end-diastolic pressure-volume relationship (EDPVR). The function $e(t)$ describes the time course of the chamber stiffness between end systole and end diastole and is defined as follows
    \begin{align} \label{eqn:timefunc}  
      e(t) = 
      \begin{cases}
        0.5 (1 - \cos \dfrac{\pi t}{T_{max, \,lv}}); &t \le t_{tr, \,lv} \\
        0.5 \exp(-(t - t_{tr,\, lv})\dfrac{1}{\tau_{lv}}); &t_{tr, \,lv} < t \le T_{c}\,,
      \end{cases}
    \end{align}
    where $t$ is the time elapsed since the beginning of systole, $T_{max,\, lv}$ is the time-to-peak-tension,  $t_{tr, lv}$ is the transition time,   $\tau_{lv}$ is the relaxation time constant, and $T_c$ denotes the duration of a cardiac cycle.
    
    The same relations hold for the left atrium with a time-shift, $t_{la}$, that is given by
    \begin{align}   
      t_{la} = 
      \begin{cases}
        t + 0.1 T_{c}; &t \le 0.9 T_{c} \\
        t - 0.9 T_{c}; &t > 0.9 T_{c}\,.
      \end{cases}
    \end{align}

 \noindent In our model, the end-systolic elastance $E_{es,\, la}$ and transition time $t_{tr, ,la}$ of the left atrium are held constant, while the left ventricle's contractility $E_{es,\, lv}$ and transient time $t_{tr,\, lv}$ are treated as model inputs. To ease the notation, we use $E_{es}$ and $t_{tr}$ (instead of $E_{es,\, lv}$, and $t_{tr, \,lv}$) to denote the contractility and transient time of the left ventricle throughout the rest of this paper.

  \subsection{Physics-informed neural network (PINN)}
    In addition to time $t$, the PINN model incorporates 10 model parameters as inputs, i.e., $R_{av}$, $R_{ao}$, $C_{ao}$, $R_{art}$, $C_{art}$, $R_{vc}$, $C_{vc}$, $R_{mv}$, $E_{es}$, and $t_{tr}$. The baseline value of these parameters are derived from previous studies \cite{Danielsen, Mller2014AGM, Shavik2020}. These values are modified using a set of multipliers, which serves as the actual model inputs. Here, the allowable range for the multipliers is $[0.3,\, 3.0]$ except for the multiplier of $t_{tr}$, which has a range $[0.8,\, 1.2]$. Details of the baseline values and model constants are provided in \autoref{tab:baseline} in the Appendix \ref{appendix:MP}. We note that while selecting a larger input range for the multipliers can enhance the model's ability to capture a wider range of outputs, it will also complicate the training of the PINN. 

\begin{figure}[ht]\centering
      \includegraphics[width=0.9\textwidth]{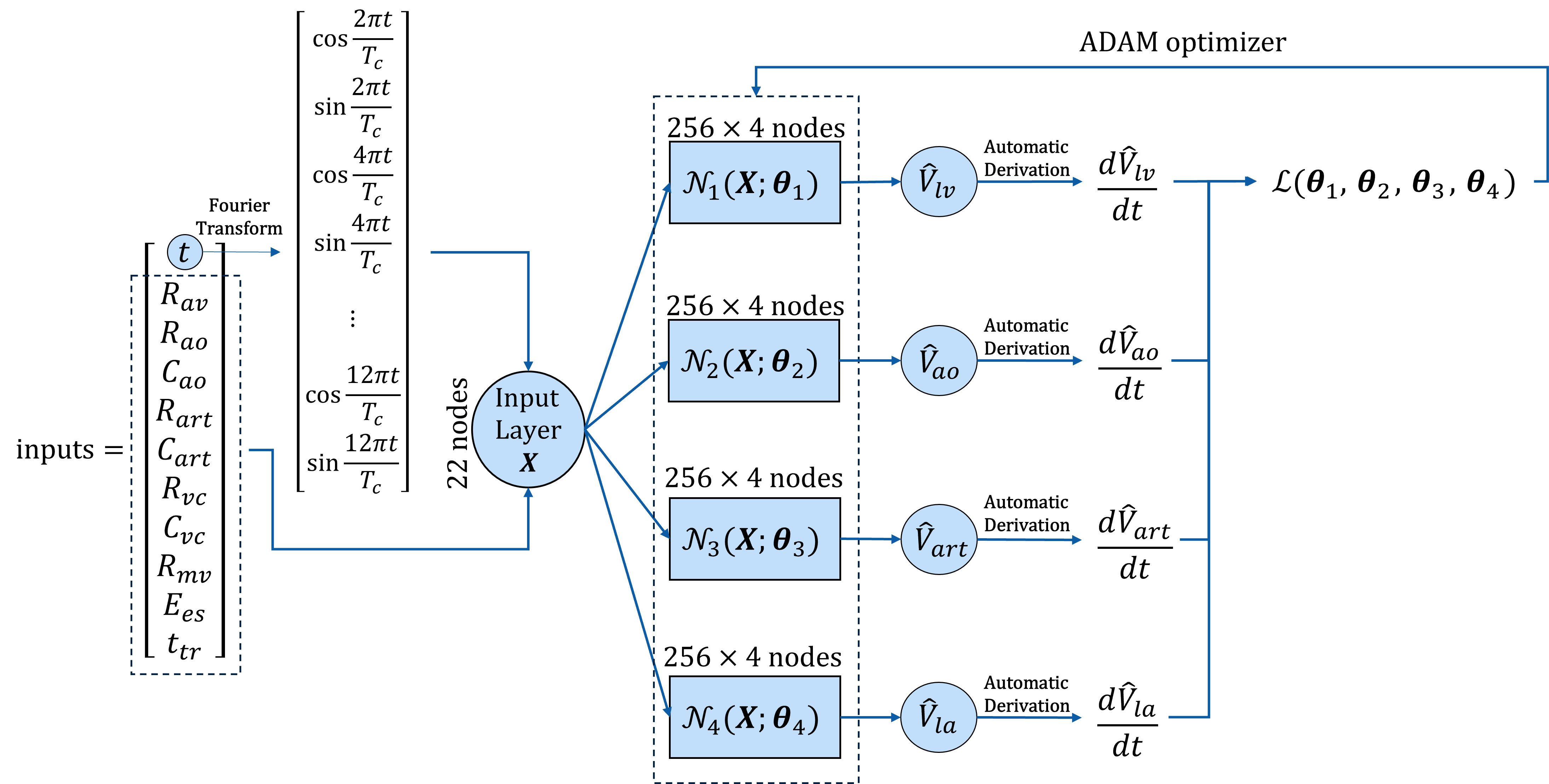} 
      \caption{Model architecture: Input layer size is 22, comprising 12 Fourier terms and 10 input parameters. Four separate neural networks, each characterized by its set of weights and biases represented as $\boldsymbol{\theta}_{i}$.}\label{fig:nn}
    \end{figure}

    Four separate neural networks, each with a size of $256 \times 4$, are used here. A schematic of the network architecture is shown in \autoref{fig:nn}. The output of each network represents the volume of one compartment of the closed loop system (\autoref{fig:Circuit}). 
    The volume of the fifth component (volume of the vena cava) is calculated based on mass conservation, i.e.,
    \begin{align}   
      \hat{V}_{vc} = V_{total} - (\hat{V}_{lv} + \hat{V}_{ao} + \hat{V}_{art} + \hat{V}_{la}).
    \end{align}
\noindent We note here that PINN-based model predictions are marked with the `hat' symbol (e.g., $\hat{V}_{vc}$) to distinguish them from the true values (e.g., $V_{vc}$).

    To enforce the periodic boundary conditions (steady state solution), the volume waveforms are each represented by the first twelve terms of the Fourier series suggested in \cite{Dong_2021, Lu_BC_2021}. The activation function $\tanh$ is used in all layers except the output layer.

    A dataset comprising \num{100000} cases was generated for the training phase using Latin Hypercube Sampling (LHS) \cite{LHCS} to sample the parameter space. Each case represents a unique combination of input parameters. An additional \num{100000} cases were selected using the same LHS method to create a test set. Each case has \num{400} timepoints. In total, we have \num{40000000} collocation points for training and another \num{40000000} collocation points for testing.

    The optimization problem for training the PINN model is defined as follows
    \begin{align}\label{eqn:optimization}
      \boldsymbol{\theta}^* = \argmin_{\boldsymbol{\theta}} \mathcal{L}(\boldsymbol{\theta}) \hspace{1pt},
    \end{align}
    where $\mathcal{L}$ is the loss function, $\boldsymbol{\theta}$ denotes $\left[\boldsymbol{\theta}_{1}^T, \boldsymbol{\theta}_{2}^T, \boldsymbol{\theta}_{3}^T, \boldsymbol{\theta}_{4}^T \right]^T$, and $\boldsymbol{\theta}_{1}$, $\boldsymbol{\theta}_{2}$, $\boldsymbol{\theta}_{3}$ and $\boldsymbol{\theta}_{4}$ are the vectors of all the weights and biases for their respective networks. 

    The PINN model is implemented using PyTorch. The derivatives of the volume variables ($\hat{V}_{lv}, \hat{V}_{ao}, \hat{V}_{art}, \hat{V}_{vc}, \hat{V}_{la}$) with respect to time are calculated using the Automatic Derivation \cite{NEURIPS2019_9015}. To solve the optimization problem (\autoref{eqn:optimization}), the builtin ADAM optimizer was utilized with an initial learning rate of $10^{-3}$ \cite{Adam2014}. The learning rate was dynamically adjusted during training. Specifically, if the loss stops decreasing, the learning rate was reduced by a factor of two. The minimum learning rate was set to $10^{-6}$.
    
    To stabilize the training and prevent the model's hyperparameters from being trapped in sharp local minima, the following strategies were used:
    \begin{enumerate}
      \item  The inputs were scaled between 0 to 1 (a common practice) \cite{Aggarwal2018} whereas the scaling range for the outputs was adjusted to -1 and 1 that aligns with the range of the activation function $\tanh$.

      \item  The Mean Absolute Error (MAE) was used as the loss function to start the training. Once the learning rate becomes smaller than $10^{-4}$, the loss function was switched to the Mean Square Error (MSE). The MSE loss function is defined as the average of squared residuals for the five ODEs (Eq.~\ref{eqn:ODEs}) evaluated at all the collocation points, i.e.,
      \begin{equation}\label{eqn:loss}
        \begin{split}
          \mathcal{L}_{MSE}(\boldsymbol{\theta})
          = \dfrac{1}{\mathcal{N}_{c}}\sum_{i=1}^{\mathcal{N}_{c}} &\left(
            \left[\frac{d\hat{V}_{lv}}{dt} - (q_{mv} - q_{av}) \right]_i^2 +
            \left[\frac{d\hat{V}_{ao}}{dt} - (q_{av} - q_{ao}) \right]_i^2 \right. \\
            &+ \left[\frac{d\hat{V}_{art}}{dt} - (q_{ao} - q_{art}) \right]_i^2 +
            \left[\frac{d\hat{V}_{vc}}{dt} - (q_{art} - q_{vc}) \right]_i^2 \\
            &\left. + \left[\frac{d\hat{V}_{la}}{dt} - (q_{vc} - q_{mv}) \right]_i^2 \right),
        \end{split}
        \end{equation}
      where $\mathcal{N}_{c}$ denotes the number of collocation points. Similarly, the loss function based on MAE is defined as
      \begin{equation}\label{eqn:lossMAE}
        \begin{split}
          \mathcal{L}_{MAE}(\boldsymbol{\theta})
          = \dfrac{1}{\mathcal{N}_{c}}\sum_{i=1}^{\mathcal{N}_{c}} &\left(
            \left|\frac{d\hat{V}_{lv}}{dt} - (q_{mv} - q_{av}) \right|_i +
            \left|\frac{d\hat{V}_{ao}}{dt} - (q_{av} - q_{ao}) \right|_i \right. \\
            &+ \left|\frac{d\hat{V}_{art}}{dt} - (q_{ao} - q_{art}) \right|_i +
            \left|\frac{d\hat{V}_{vc}}{dt} - (q_{art} - q_{vc}) \right|_i\\
            &\left. + \left|\frac{d\hat{V}_{la}}{dt} - (q_{vc} - q_{mv}) \right|_i \right).
        \end{split}
      \end{equation}

      This approach produced better results than continuing the training with MAE or starting the training with MSE as the loss function because both approaches often resulted in the training algorithm being trapped in a local minima.

      \item  A small $\alpha$ (with a value of $0.002$) was used in the smooth maximum function (\autoref{eqn:smoothmax}) to avoid the training algorithm from being trapped in the local minima at the beginning. When the learning rate becomes smaller than $10^{-4}$, the value of $\alpha$ was adjusted to get a closer approximation of the maximum function (i.e., $\alpha=0.01$).
    \end{enumerate}

\noindent To evaluate the performance and accuracy of the trained PINN model, a separate set of \num{1000} cases was generated to compare the model results with the numerical solution of ODEs that were solved using the forward Euler Method. The evaluation metric employed here is the Relative Mean Absolute Error (RMAE), as defined by
\begin{align} \label{eqn:eval_metric}
  \text{RMAE}(\hat{y}) = \dfrac{1}{N}\sum_{i=0}^{N}\left(\dfrac{|\hat{y} - y_{N}|}{y_{N}}\right)_i  \hspace{2pt},
\end{align}

\noindent where $y_{N}$ denotes the numerical solution. The model was trained on four Nvidia Tesla V100 GPUs in the High-performance Cloud Computing nodes at Michigan State University.

  \subsection{Sensitivity analysis}
   Sobol sensitivity analysis was performed to assess the influence of each model input parameter on the model outputs. The analysis decomposes the total variance in the model output into contributions from individual model input parameters and their mutual interactions \cite{SOBOL2001, SALTELLI2002, SALTELLI2010, CAMPOLONGO2011978, Owen2020OnDT}. To outline the mathematical framework used to calculate the Sobol indices, we denote the PINN model as a function $f$ of the input parameters $\mathbf{x} = (x_1, x_2, \dots, x_s)$ that are rescaled to the unit interval $[0, 1]$.  Each parameter can be conceptualized as a random variable following a uniform distribution, with the assumption of mutually independence. In this probabilistic interpretation, the function $f$ is characterized as a random variable with mean $f_0$ and variance $V$ defined as 
   \begin{align}
      f_0 &= \int_{\Omega} f(\mathbf{x}) \prod_{i=1}^{s} dx_i \,, \\
      V &= \int_{\Omega} f^2(\mathbf{x}) \prod_{i=1}^{s} dx_i - f_0^2 \,.
   \end{align}

\noindent The Sobol method decomposes the total variance $V$ into summands representing contributions from individual parameters, including their interactions with each other. This is done by decomposing $f(\mathbf{x})$ into
   \begin{align} \label{eqn:sobol_decopmosition}
      f(\mathbf{x}) = f_0 + \sum_{i=1}^{s} f_i(x_i) + \dots + \sum_{i=1}^{s} \sum_{j \ne i}^{s}f_{ij}(x_i, \, x_j) + \dots + f_{12 \dots s}(\mathbf{x}) \,,
   \end{align}
   where the decomposition terms can be calculated by following relations
   \begin{align}
      f_i(x_i) = &\int_{\Omega_{\sim i}} f(\mathbf{x}) \prod_{k \ne i} dx_k - f_0 \,, \\
      f_{ij}(x_i, \, x_j) = &\int_{\Omega_{\sim(i, j)}} f(\mathbf{x}) \prod_{k \ne i, j} dx_k - f_i(x_i) - f_j(x_j) - f_0 \,,
   \end{align}
   By $\Omega_{\sim i}$, we denote the parameter space of all components except $x_i$. The higher order terms are calculated in a similar way. It can be shown that all the summands are orthogonal, which yields
   \begin{align}
      V = \sum_{i=1}^{s} V_i + \sum_{i<j} V_{ij} + \sum_{i < j < k} V_{ijk} + ... + V_{12 \dots s} \,,
   \end{align}
   where
   \begin{align}
      V_i = \int_{\Omega} f_i^2 \prod_{k=1}^{s} dx_k \,.
   \end{align}

\noindent The Sobol sensitivity indices are defined as
   \begin{align}
      S_i = \dfrac{V_i}{V} \,,
   \end{align}
\noindent and the total order Sobol indices are defined as
   \begin{align}
      S_{Ti} = S_i + S_{ij} + ... + S_{1 \dots i \dots s} \,.
   \end{align}

\noindent Here, only the first and second order Sobol indices were calculated, and the total order Sobol indices are the summation of the first and second order indices. The Sobol integrals were evaluated using the Monte Carlo Method with \num{360000} functional evaluations. Time-average value of the indices, which were computed over \num{200} timepoints is reported here. The SALib package in Python \cite{Herman2017, Iwanaga2022} was used to compute the Sobol indices. 

  \subsection{Inverse modeling}
    Once the PINN model was trained, it was then used for inverse modeling to estimate parameters from experimental measurements.

  \subsubsection{Experimental data}
    To validate the model, experimental data were acquired from 11 Yorkshire domestic swine. Left ventricular (LV) pressure waveform was acquired in each animal with a pressure catheter that was inserted through the carotid or femoral artery \cite{FanRoleofCoro2021}. Left ventricular volume waveform was obtained from 3D ECHO images. These images were acquired using an EPIQ 7C ultrasound system (Philips, Andover, MA) with the animal in a supine position \cite{FAN2022105050}. In 3 animals, the waveforms were acquired both before and after the administration of dobutamine, an inotropic agent that increases the myocardial contractility \cite{Tuttle1975, burkhoff1987influence, martens2018effects}. The recorded LV pressure and volume waveforms were used in the inverse modeling process to estimate model parameters and to test whether this process can quantify the increase in contractility associated with the administration of dobutamine.
All animal experiments were performed in accordance with national and local ethical guidelines, including the Guide for the Care and Use of Laboratory Animals, the Public Health Service Policy on Humane Care and Use of Laboratory Animals, and the Animal Welfare Act, and an approved California Medical Innovations Institute IACUC protocol regarding the use of animals in research.

  \subsubsection{Parameter estimation}
   The goal here is to use given single-beat LV pressure and volume waveforms to estimate an optimal combination of  input parameters that maximizes the coefficient of determination ($R^2$) between the PINN model predicted waveforms and the given waveforms. The coefficient of determination is defined by:
 
    \begin{align} \label{eqn:R2}
      R^2 = 1 - \dfrac{\sum_{i=0}^{N}(y_{true} - \hat{y})_i^2}{\sum_{i=0}^{N}(y_{true} - \bar{y})_i^2}\,,
    \end{align}
    where $\bar{y}$ is the average of known data, i.e.,
    \begin{align*}
      \bar{y}=\frac{1}{N} \sum_{i=0}^{N}y_{true_i}\,.
    \end{align*}
    A value of 0 for $R^2$ indicates poor accuracy of the regression (akin to a simple mean-based model), while a value of 1 indicates that the model's predictions capture the data perfectly. In essence, $R^2$ serves as a metric to evaluate how well a regression model aligns with the actual data, with higher values indicating a closer fit.

    The differential evolution method \cite{Storn1997, Price2005} with a population size of \num{100} and a maximum of \num{200} generations was used to perform the optimization. The parameter intervals employed during optimization are within the limits used for training the PINN model. The optimization was executed using SciPy \cite{2020SciPy-NMeth}.

    To investigate the uniqueness of the model parameters, parameter estimation was also performed on \num{100} pairs of synthetic waveforms generated by the PINN model with distinct sets of parameter values to determine whether the same values used for generating the waveforms are recovered with inverse modeling.

\section{Results}

\subsection{Training of the PINN}

\begin{figure}[ht]
   \centering
    \begin{subfigure}{0.4\textwidth}
          \includegraphics[width=\linewidth]{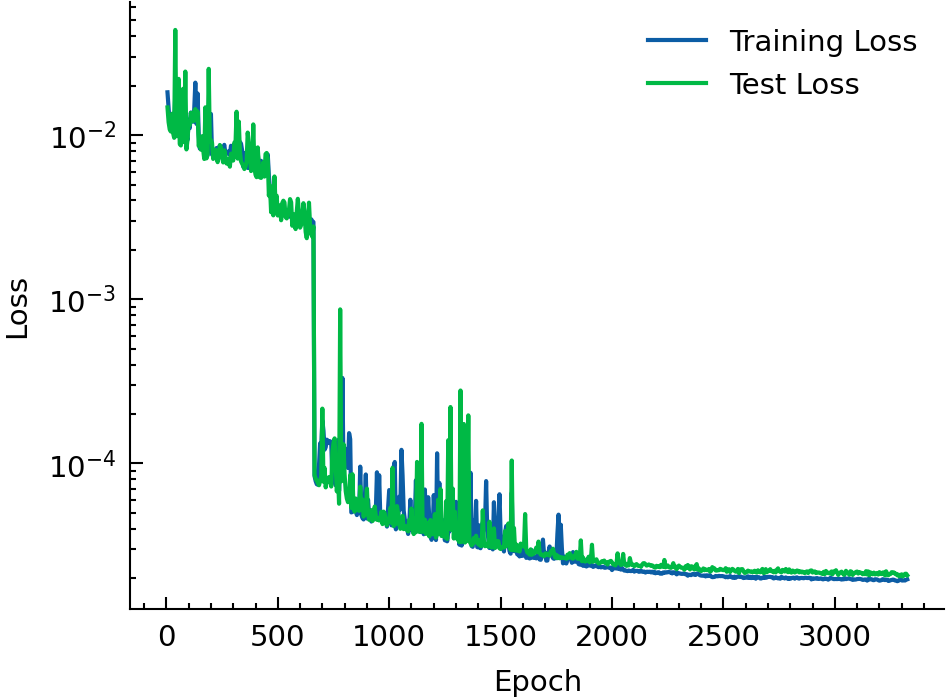}
          \caption{Training and test loss curves.}
          \label{fig:losses}
     \end{subfigure} 
        \hfill
    \begin{subfigure}{0.4\textwidth}
          \includegraphics[width=\linewidth]{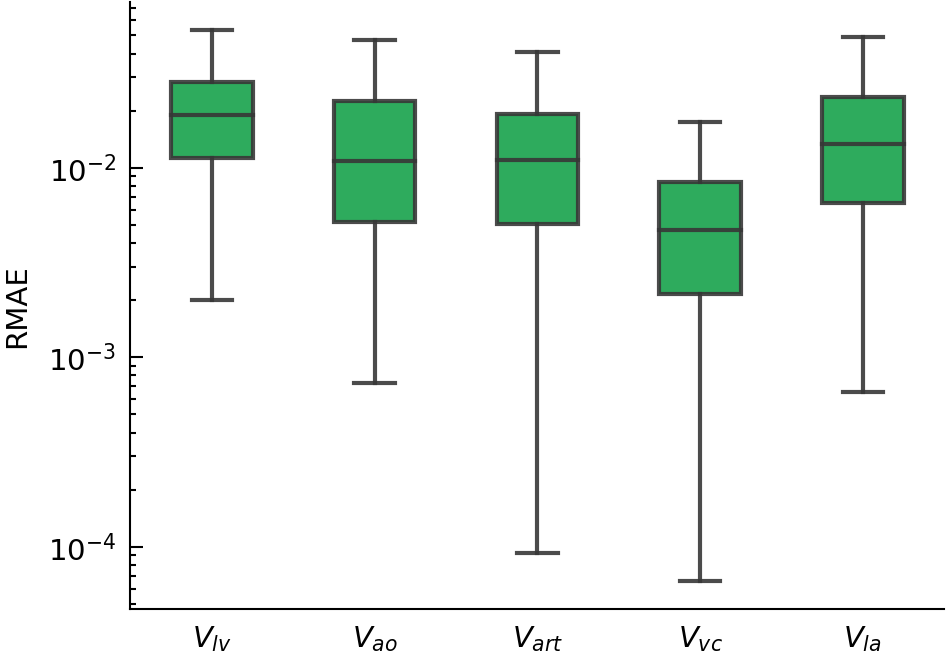}
          \caption{RMAE between the PINN model and numerical results in the \num{1000} evalutaion cases}\label{fig:pinn_error}
    \end{subfigure}
    \caption{Results of the PINN training}
\end{figure}

\noindent \autoref{fig:losses} shows the training and test losses of the PINN model, which was trained for a total of \num{3300} epochs. The noticeable decrease in losses around epoch \num{600} corresponds the transition from using the MAE to using the MSE in the loss function.
\autoref{fig:pinn_error} shows the RMAE between the PINN model and numerical results of the 5 volume state variables (\autoref{eqn:ODEs}) in the \num{1000} evaluation cases. Average of the errors are all below $2\%$, with the largest error not exceeding $5\%$. The smallest error is found in the volume associated with the vena cava.

\subsection{Sensitivity analysis}\label{subsec:R_SA}
\begin{figure}[ht]\centering
  \includegraphics[width=0.8\textwidth]{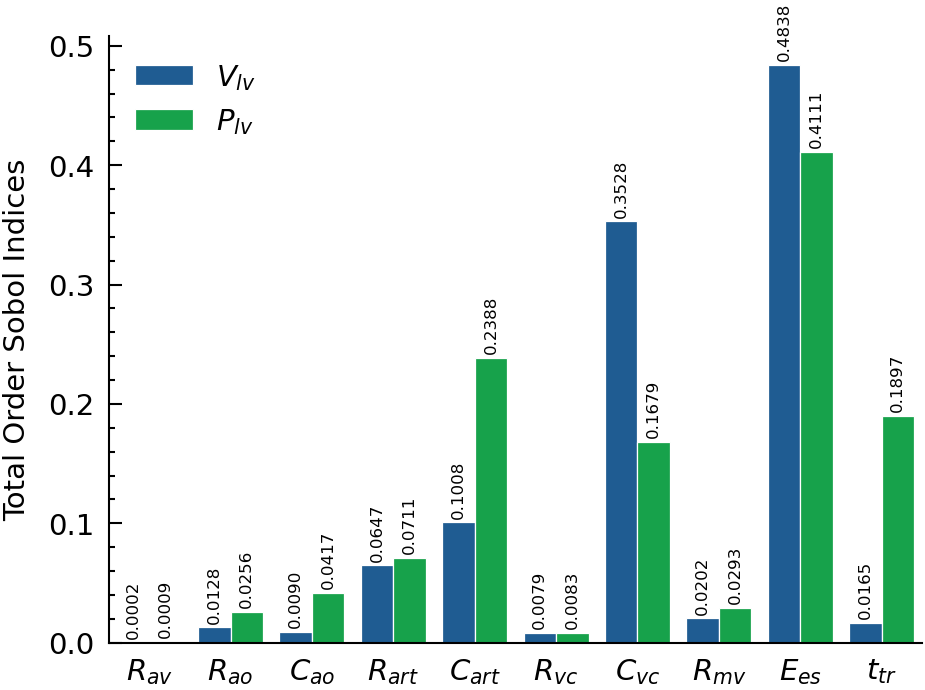}
  \caption{Total order Sobol indices associated with $V_{lv}$ and $P_{lv}$ for each input parameters}\label{fig:SA_VPT}
\end{figure}

\autoref{fig:SA_VPT} shows the total order Sobol indices associated with $V_{lv}$ and $P_{lv}$ for each input parameter. The LV volume $V_{lv}$ is most sensitive to contractility $E_{es}$  followed (in order) by the vena cava compliance $C_{vc}$, peripheral artery compliance $C_{art}$, and its resistance $R_{art}$. The LV pressure waveform is also most sensitive to $E_{es}$, and is followed by $C_{art}$ and the transition time $t_{tr}$. Both $V_{lv}$ and $P_{lv}$ are insensitive to $R_{av}$, $R_{ao}$, $C_{ao}$, $R_{vc}$, and $R_{mv}$ within the parameter range applied in this study, with Sobol indices below $0.05$.

\subsection{Inverse modeling on synthetic data}
\begin{figure}[htbp]\centering
  \includegraphics[width=0.8\textwidth]{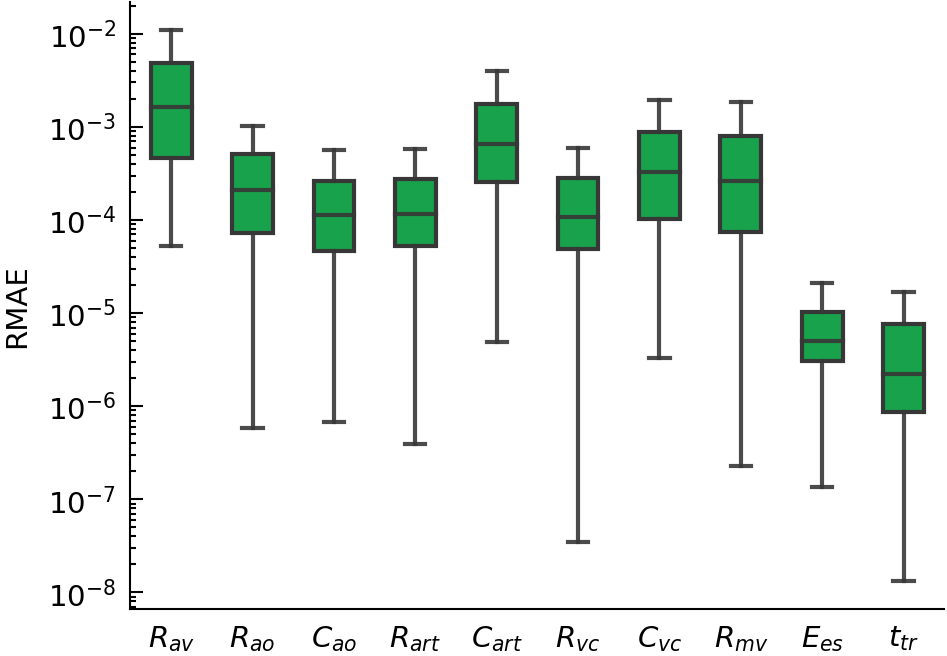} 
  \caption{RMAE between the estimated input parameters and their ground truth for the \num{100} synthetic cases}\label{fig:InvUniqueness}
\end{figure}

\autoref{fig:InvUniqueness} shows the RMAE between parameters estimated based on the 100 pairs of synthetic waveforms generated by the PINN model and their ground truth. The RMAE is below $1\%$ for all parameters. We also note that the inverse model completely recovers the original waveform, where the fitted waveforms all have an $R^2$ exceeding \num{0.998}.

\subsection{Inverse modeling on experimental measurements}
\begin{figure}[ht]\centering
  \begin{subfigure}[b]{0.98\textwidth}
    \includegraphics[width=\textwidth]{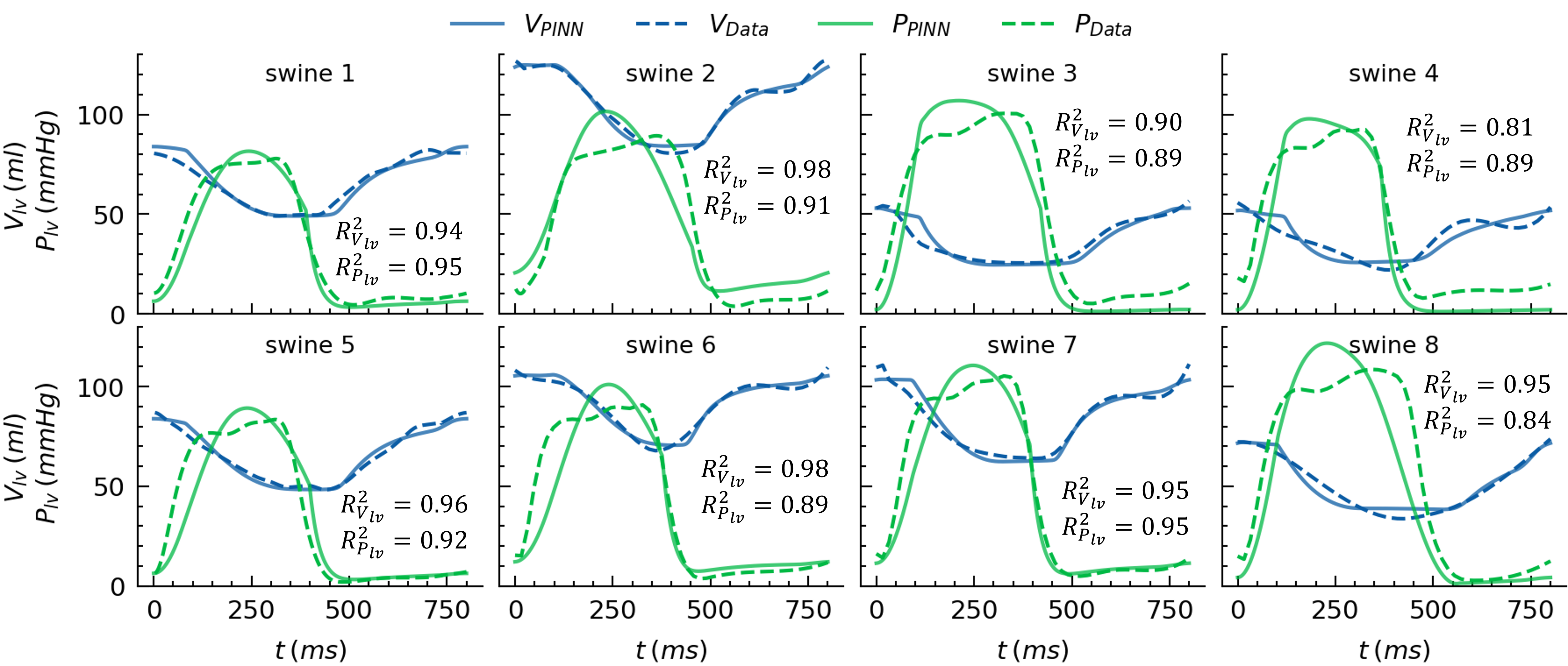}
    \caption{Eight swine subjects at baseline}
    \label{fig:InvModelingBase}
  \end{subfigure}

  \begin{subfigure}[b]{0.85\textwidth}
    \includegraphics[width=\textwidth]{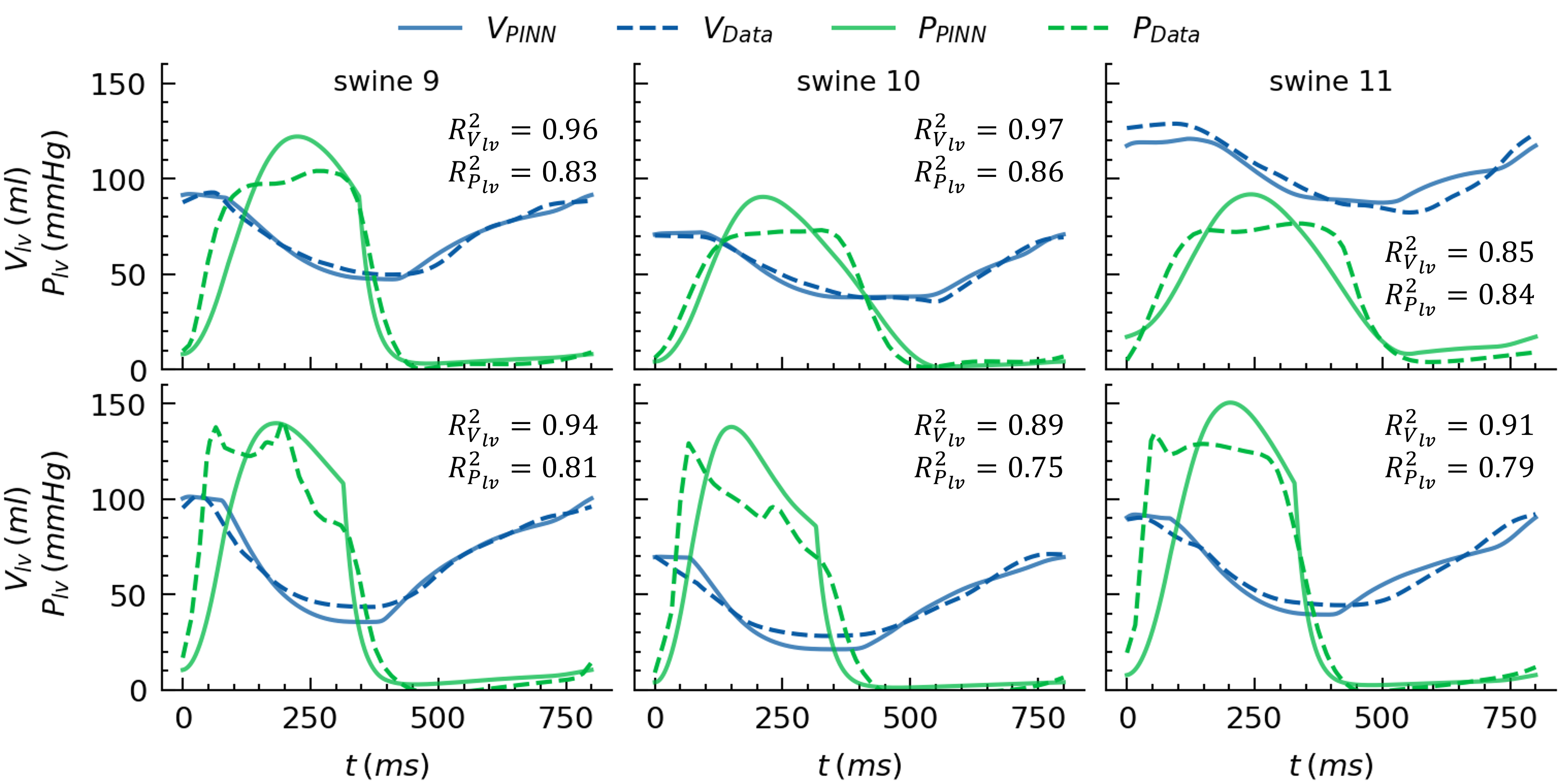}
    \caption{Three swine subjects before (top panel) and after (bottom) administration of dobutamine }
    \label{fig:InvModelingBasevsDbtmn}
  \end{subfigure}
  \caption{Comparison of the fitted LV pressure and volume waveforms from inverse modeling with the corresponding measurements from the swine models}
  \label{fig:InvModeling}
\end{figure}

\autoref{fig:InvModeling} shows the results of inverse modeling performed using the trained PINN model with measurements from 11 swine model, including measurements acquired in 3 swine model  before and after the administration of dobutamine. The fitted LV pressure and volume waveforms agree well with the experimental measurements, as reflected by the large value of $R^2$. Specifically, the lowest $R^2$ value for the LV volume waveforms and the LV pressure waveforms is \num{0.81} and \num{0.75}, respectively.

\autoref{tab:InvModeling} shows the parameter values estimated from inverse modeling performed on the data of each swine. For the last three swine, for which measurements are available before and after dobutamine administration, the estimated $E_{es}$ (i.e., contractility) associated with dobutamine is higher. Specifically, $E_{es}$ was estimated to increase from $2.73$, $2.50$, and $1.00$ to $4.34$, $7.82$, and $3.84$, respectively, in those animals after dobutamine was administered. The transient time ($t_{tr}$) estimated by the PINN model with measurements after the administration of dobutamine is also lower compared to the values estimated with measurements before the administration.

\begin{table}
  \centering
  \caption{Estimated parameters for each set of swine data. Units for resistances, compliances, LV contractility  ($E_{es}$) and transition time ($t_{tr})$ are $mmHg\: ms/ml$,  $ml/mmHg$, $mmHg/ml$ and $ms$, respectively.}
  \label{tab:InvModeling}
  \begin{tabular*}{\textwidth}{@{\extracolsep{\fill}}cccccccccccc}
    \toprule
    \multirow{2}{*}{\makecell{Animal}} & \multirow{2}{*}{State} & \multirow{2}{*}{$R_{av}$} & \multirow{2}{*}{$R_{ao}$} & \multirow{2}{*}{$C_{ao}$} & \multirow{2}{*}{$R_{art}$} & \multirow{2}{*}{$C_{art}$} & \multirow{2}{*}{$R_{vc}$} & \multirow{2}{*}{$C_{vc}$} & \multirow{2}{*}{$R_{mv}$} & \multirow{2}{*}{$E_{es}$} & \multirow{2}{*}{$t_{tr}$} \\
      & & & & & & & & & & & \\
    \midrule
    1 & \small{baseline} & 1.8 & 720.0 & 0.3 & 389.2 & 1.7 & 2.7 & 370.4 & 1.2 & 1.96 & 389 \\
    2 & \small{baseline} & 11.7 & 241.9 & 0.3 & 655.7 & 10.0 & 19.6 & 42.8 & 1.2 & 1.22 & 453 \\
    3 & \small{baseline} & 1.8 & 306.6 & 1.0 & 3167.0 & 10.0 & 2.7 & 400.0 & 1.2 & 7.09 & 420 \\
    4 & \small{baseline} & 1.8 & 77.4 & 1.0 & 3375.0 & 10.0 & 13.7 & 400.0 & 1.2 & 5.82 & 366 \\
    5 & \small{baseline} & 1.8 & 720.0 & 0.2 & 353.8 & 3.9 & 3.3 & 336.4 & 1.5 & 2.10 & 398 \\
    6 & \small{baseline} & 18.0 & 382.2 & 0.2 & 546.5 & 7.6 & 2.7 & 114.8 & 1.2 & 1.44 & 376 \\
    7 & \small{baseline} & 1.8 & 720.0 & 0.4 & 563.4 & 8.0 & 4.7 & 125.3 & 1.2 & 2.03 & 393 \\
    8 & \small{baseline} & 1.8 & 561.5 & 0.4 & 1862.5 & 1.0 & 5.6 & 400.0 & 4.1 & 3.98 & 567 \\
    \cmidrule(lr){1-12}
    \multirow{2}{*}{9} & \small{baseline} & 5.2 & 436.6 & 0.2 & 817.1 & 1.0 & 7.9 & 181.0 & 12.4 & 2.73 & 345 \\
    & \footnotesize{dobutamine} & 11.8 & 210.2 & 0.4 & 865.2 & 1.0 & 14.7 & 128.4 & 12.4 & 4.34 & 314 \\
    \cmidrule(lr){2-12}
    \multirow{2}{*}{10} & \small{baseline} & 18.0 & 341.9 & 0.1 & 834.4 & 10.0 & 3.0 & 274.8 & 7.4 & 2.50 & 567 \\
    & \footnotesize{dobutamine} & 1.8 & 213.9 & 0.1 & 1111.9 & 1.0 & 2.7 & 316.1 & 12.4 & 7.82 & 316 \\
    \cmidrule(lr){2-12}
    \multirow{2}{*}{11} & \small{baseline} & 18.0 & 278.9 & 0.1 & 728.8 & 10.0 & 27.0 & 59.4 & 5.8 & 1.00 & 567 \\
    & \footnotesize{dobutamine} & 1.8 & 322.6 & 0.1 & 1092.9 & 1.0 & 27.0 & 162.2 & 12.4 & 3.84 & 328 \\
    \bottomrule
  \end{tabular*}
\end{table}

\section{Discussion}

We have developed and trained a PINN model that is based on the mathematical ODE formulation of a closed-loop lumped parameter description of the cardiovascular system. The trained PINN model can be used for inverse modeling to estimate parameters reflecting the heart function in less than 3 minutes. We also validated the inverse PINN modeling approach using experimental measurements acquired from a swine model with and without the administration of an inotropic agent.

\subsection{Accuracy of the PINN model}
The training strategy we adopted here enables us to minimize the loss function, resulting in an error averaging less than $2\%$ across the entire parameter space for all five volume state variables when compared to the numerical model. The largest observed error is $5\%$, which is within the level of accuracy found in other studies using PINN. 
Specifically,  a maximum error of around $10\%$ and an average error of less than $2\%$ are reported for a PINN model developed to estimate material properties and predict the deformation  of  aorta  \cite{LIANG2023116347}. Similarly, errors below $1\%$ and $7\%$ are reported for PINN models developed to simulate soft-tissue mechanics \cite{DALTON2023116351} and tube flows based on the Navier-Stokes equation \cite{Wong2023}. We note here that the number of parameters used as model input in these studies (between 2 to 5 parameters) is much smaller compared to that (10 parameters) used here.

\subsection{Computational time}
Training of the PINN model takes 96 hours. However, the trained PINN model can be used to obtain the Sobol sensitivity analysis, which requires \num{360000} model evaluations, within \num{30} minutes. Furthermore, inverse modeling for each case, which may require up to \num{20000} model evalutions, is completed in less than 3 minutes. We note that given that solving the ODEs numerically takes around 10 seconds to obtain the steady-state solution, performing the Sobol sensitivity analysis and the inverse modeling will require 1000 (cf. \num{30} mins) and 5.56 hours (cf. \num{3} mins), respectively. These results demonstrate the ability of the PINN model to efficiently perform tasks that would otherwise be significantly time-consuming if numerical methods are used.

\subsection{Sensitivity analysis of the PINN model parameters}
The PINN model enables us to perform a global sensitivity study efficiently where we found that $V_{lv}$ is most sensitive to $E_{es}$, followed by $C_{vc}$, $C_{art}$ and $R_{art}$  (\autoref{fig:SA_VPT}). The LV pressure $P_{lv}$ is most sensitive to $E_{es}$, followed by $C_{art}$, $t_{tr}$, $C_{vc}$ and $R_{art}$. The finding that $E_{es}$, which reflects the LV contractility, affects the LV volume and pressure waveform the most is expected. The results where the compliance of the vena cava $C_{vc}$ and arteries $C_{art}$ also substantially affects the LV pressure and volume waveforms is also expected as a change in $C_{vc}$ and $C_{art}$ affects the preload and afterload imposed on the LV, respectively.

\subsection{Parameter estimation using the trained PINN model}
Our findings indicate that inverse modeling using the PINN model can successfully recover the generated synthetic single-beat LV volume and pressure waveforms in all the 100 evaluation cases as well as their corresponding model parameters. This result implies that these parameters are unique. Correspondingly, parameters estimated from inverse modeling using experimental measurements of the volume and pressure waveforms are likely to be also unique (\autoref{fig:InvModeling} and \autoref{tab:InvModeling}), which provide confidence of using them to reflect certain physiological and pathological conditions. The smaller error in the estimation of $E_{es}$ and $t_{tr}$ parameters suggests a higher reliability in their values compared to the rest of the parameters.

The results from inverse modeling performed on the experimental measurements acquired in the animals after dobutamine administration show an increase in the estimated $E_{es}$ from $2.73$, $2.50$, and $1.00$ at baseline to $4.34$, $7.82$, and $3.84$, respectively. 
Correspondingly, the percentage increase of the estimated $E_{es}$ associated with the administration of dobutamine ranges between 58\% to 284\%.
This range is higher than the values reported in previous studies, where the percentage increase is between $30\%$ to $50\%$ in Landrace pig \cite{Vogel2003}, canine \cite{kass1987comparative} and human \cite{leatherman1989use}. This disparity may be attributed to differences in species, dobutamine dosage and the approach by which $E_{es}$ was obtained experimentally based on manipulation of the ventricular loading conditions by occluding the vena cava \cite{Vogel2003, kass1987comparative} or pharmacologically using nitroprusside \cite{leatherman1989use}.

\subsection{Potential clinical utility} 
The rapid inverse modeling approach described here can potentially be translated clinically to estimate ventricular contractility $E_{es}$ using measurements acquired from a single heart beat (i.e., single beat estimation). Single-beat estimation of $E_{es}$ is clinically useful because it does not require multiple pressure-volume loops with different end-systolic pressure that are usually obtained by manipulating the loading conditions that is not feasible in the clinic. 
While different approaches to estimate $E_{es}$ using single-beat measurements  have been developed with reasonable accuracy \cite{takeuchi1991single,chen2001noninvasive}, they can only be used to estimate contractility but not other clinical parameters. 
The approach described here, which is grounded by physics, can potentially be used to simultaneously estimate other parameters (in addition to $E_{es}$) such as the systemic vascular resistance (reflected by $R_{art}$) or diastolic stiffness of the ventricle (as reflected by $A_{lv}$ and $B_{lv}$) using measurements acquired from a single beat.
Further investigations (e.g., determining the type of single-beat measurements), however, are needed to realize this goal.

\subsection{Limitations}
This study has some limitations. First, we only consider a limited number of model parameters to reduce the computational cost associated with training the PINN model.
Second, some of the estimated parameter values approached their upper or lower bounds (see \autoref{tab:InvModeling}), which suggests that the accuracy of inverse modeling could potentially be further improved by increasing or modifying the parameters' range to be outside of the range at which the PINN model is trained. The parameters' range used for training the PINN model can be expanded but this will increase the computational cost associated with the training. Nevertheless, we note that the PINN model predictions of $V_{lv}$ and $P_{lv}$ are not sensitive to those parameters that reach the bounds (e.g., $R_{mv}$ and $R_{av}$) (see \autoref{fig:SA_VPT}).
Last, the neural network's architecture, which was determined in an iterative process here, may not be optimal. Determining the optimal architecture and other hyperparameters of the neural network is a notable challenge  \cite{Bergstra2012281, ren2021comprehensive}. Despite significant efforts in this area \cite{wang2022, MALISIC2023}, a consistent and systematic approach for selecting an optimal network architecture  and determining optimal hyperparameters for PINNs remains elusive.

\section{Summary and conclusions}
In summary, we developed a PINN model that encodes the physics associated with the closed-loop blood circulation system consisting of the LV. 
The PINN model enables fast prediction of hemodynamics, which allows Sobol sensitivity analysis and inverse modeling to be executed rapidly. 
Parameter values of the PINN model were estimated with inverse modeling based on measurements from 11 swine models. 
The estimated end-systolic elastance $E_{es}$ is higher with measurements acquired after the administration of an inotropic agent in 3 swine model, which validates the inverse modeling approach. 
These results suggest that the PINN inverse modeling approach can potentially be further developed as a clinical tool to simultaneously estimate ventricular contractility and other physiological parameters using single-beat measurements from patients. 

\section*{Acknowledgement}
This work is supported by NIH HL160997, HL163977 and HL166508.

\section*{Code and data availability}
The code and the data are available at: \url{https://github.com/ehsanngh/lpm_pinn}

\section*{Declaration of competing interest}
The authors declare that they have no known competing financial interests or personal relationships that could have appeared to influence the work reported in this paper.

\bibliographystyle{ieeetr}
\bibliography{bibliography}

\newpage
\begin{appendices}
    \section*{Appendix}
    
\renewcommand{\thesubsection}{\Alph{subsection}}
\renewcommand{\theequation}{\thesubsection.\arabic{equation}}
\subsection{Model parameters} \label{appendix:MP}

\begin{table}[h]
    \centering
    \renewcommand{\arraystretch}{1.2}
    \caption{Constant model parameters and baseline values of the input parameters}
    \small
    \begin{tabular}{ccccc}
    \toprule
    \multicolumn{2}{c}{ } & Parameter & Value & Unit \\
    \midrule
    \multicolumn{2}{c}{\multirow{10}{*}{\makecell{Baseline values for \\input parameters}}} & $R_{av}$ & 6.0 & \multirow{5}{*}{$\mathrm{mmHg \; ms/ml}$} \\
    & & $R_{ao}$ & 240.0 & \\
    & & $R_{art}$ & 1125.0 & \\
    & & $R_{vc}$ & 9.0 & \\
    & & $R_{mv}$ & 4.1 & \\
    \cmidrule{3-5}
    & & $C_{ao}$ & 0.3 & \multirow{3}{*}{$\mathrm{ml/mmHg}$} \\
    & & $C_{art}$ & 3.0 & \\
    & & $C_{vc}$ & 133.3 & \\
    \cmidrule{3-5}
    & & $E_{es, lv}$ & 3.00 & $\mathrm{mmHg/ml}$ \\
    & & $t_{tr, lv}$ & 420 & $\mathrm{ms}$ \\
    \midrule
    \multirow{17}{*}{\makecell{Constant \\Parameters}} & \multirow{5}{*}{Closed-Loop} & $V_{total}$ & 5200 & \multirow{4}{*}{$\mathrm{ml}$} \\
    & & $V_{ao, r}$ & 100 & \\
    & & $V_{art, r}$ & 900 & \\
    & & $V_{vc, r}$ & 2800 & \\
    \cmidrule{3-5}
    & & $T_c$ & 800 & $\mathrm{ms}$ \\
    \cmidrule{2-5}
    & \multirow{5}{*}{Left-Ventricle} & $A_{lv}$ & 1.00 & $\mathrm{mmHg/ml}$ \\
    & & $B_{lv}$ & 0.027 & $1/\mathrm{ml}$ \\
    & & $V_{lv, r}$ & 10 & $\mathrm{ml}$ \\
    \cmidrule{3-5}
    & & $T_{max, lv}$ & 280 & \multirow{2}{*}{$\mathrm{ms}$} \\
    & & $\tau_{lv}$ & 25 & \\
    \cmidrule{2-5}
    & \multirow{7}{*}{Left-Atrium} & $E_{es, la}$ & 0.45 & $\mathrm{mmHg/ml}$ \\
    & & $A_{la}$ & 0.45 & $\mathrm{mmHg/ml}$ \\
    & & $B_{la}$ & 0.05 & $1/\mathrm{ml}$ \\
    & & $V_{la, r}$ & 10 & $\mathrm{ml}$ \\
    \cmidrule{3-5}
    & & $T_{max, la}$ & 150 & \multirow{3}{*}{$\mathrm{ms}$} \\
    & & $\tau_{la}$ & 25 & \\
    & & $t_{tr, la}$ & 225 & \\
    \bottomrule
    \end{tabular}
    \label{tab:baseline}
\end{table}

\clearpage

\subsection{Supplementary plots}\label{appendix:additional_plots}
\begin{figure}[ht]\centering
    \includegraphics[width=0.8\textwidth]{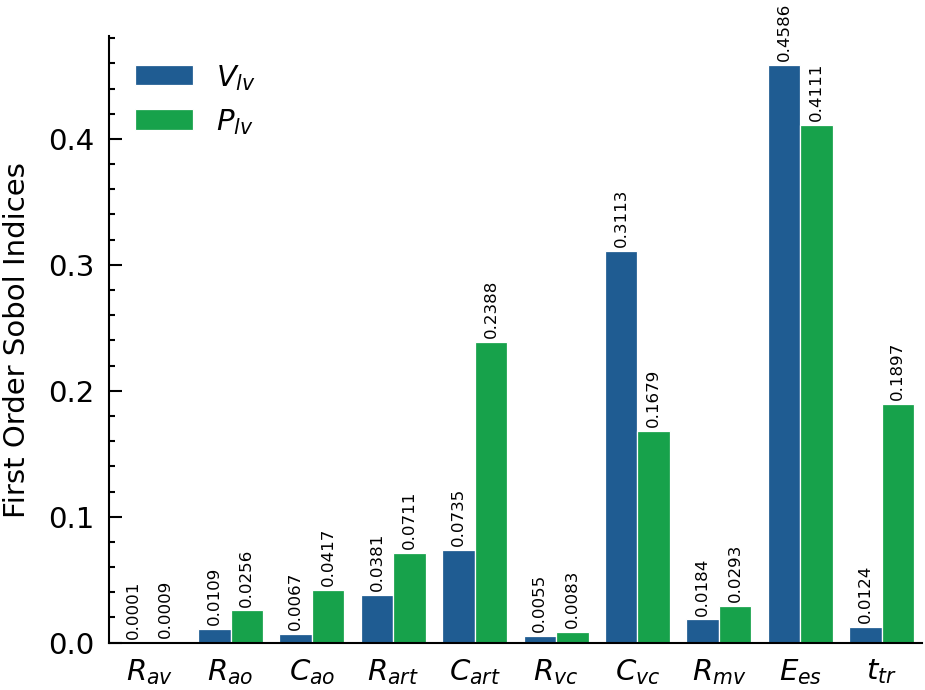} 
    \caption{First order Sobol indices associated with $V_{lv}$ and $P_{lv}$ for each input parameter}\label{fig:SA_VP1}
\end{figure}

\begin{figure}[ht]\centering
    \includegraphics[width=0.9\textwidth]{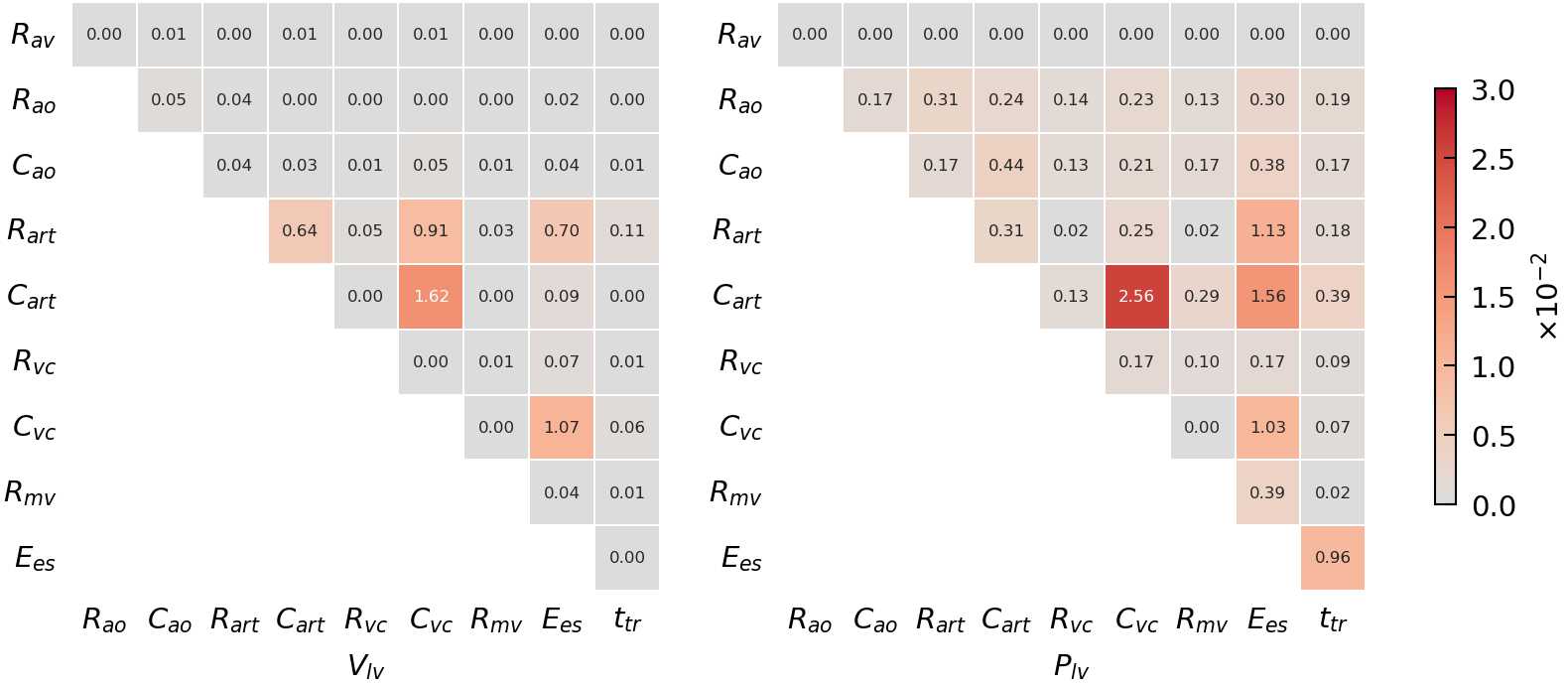}
    \caption{Second order Sobol indices associated with $V_{lv}$ and $P_{lv}$ for each input parameter}\label{fig:SA_VP2}
\end{figure}

\begin{figure}[ht]\centering
    \includegraphics[width=0.9\textwidth]{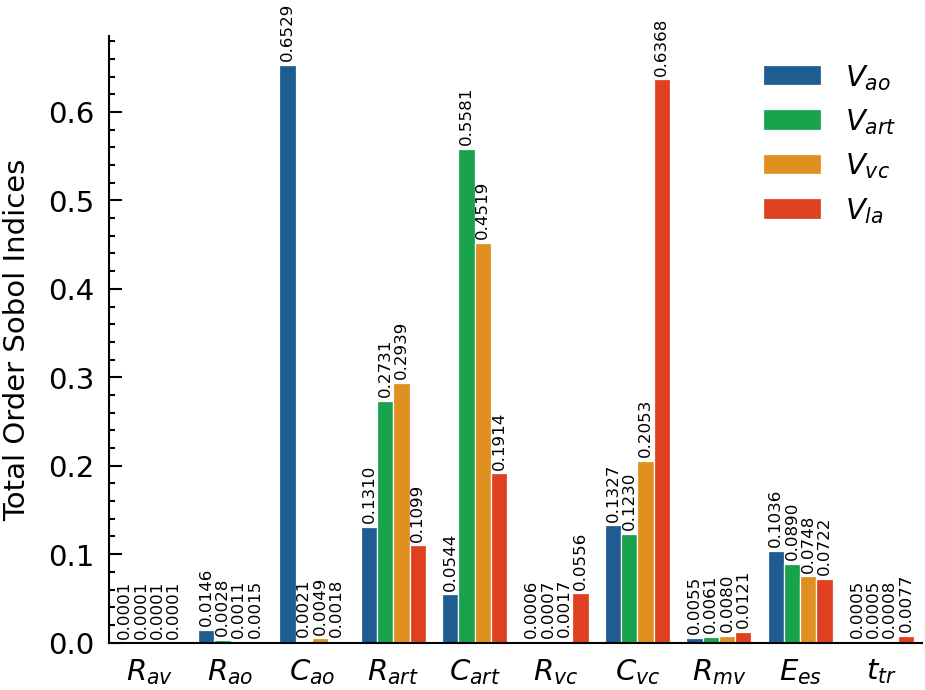}
    \caption{Total order Sobol indices associated with $V_{ao}$, $V_{art}$, $V_{vc}$, and $V_{la}$ for each input parameter}\label{fig:SA_VsT}
\end{figure}

\begin{figure}[ht]\centering
    \begin{subfigure}[b]{0.98\textwidth}
      \includegraphics[width=\textwidth]{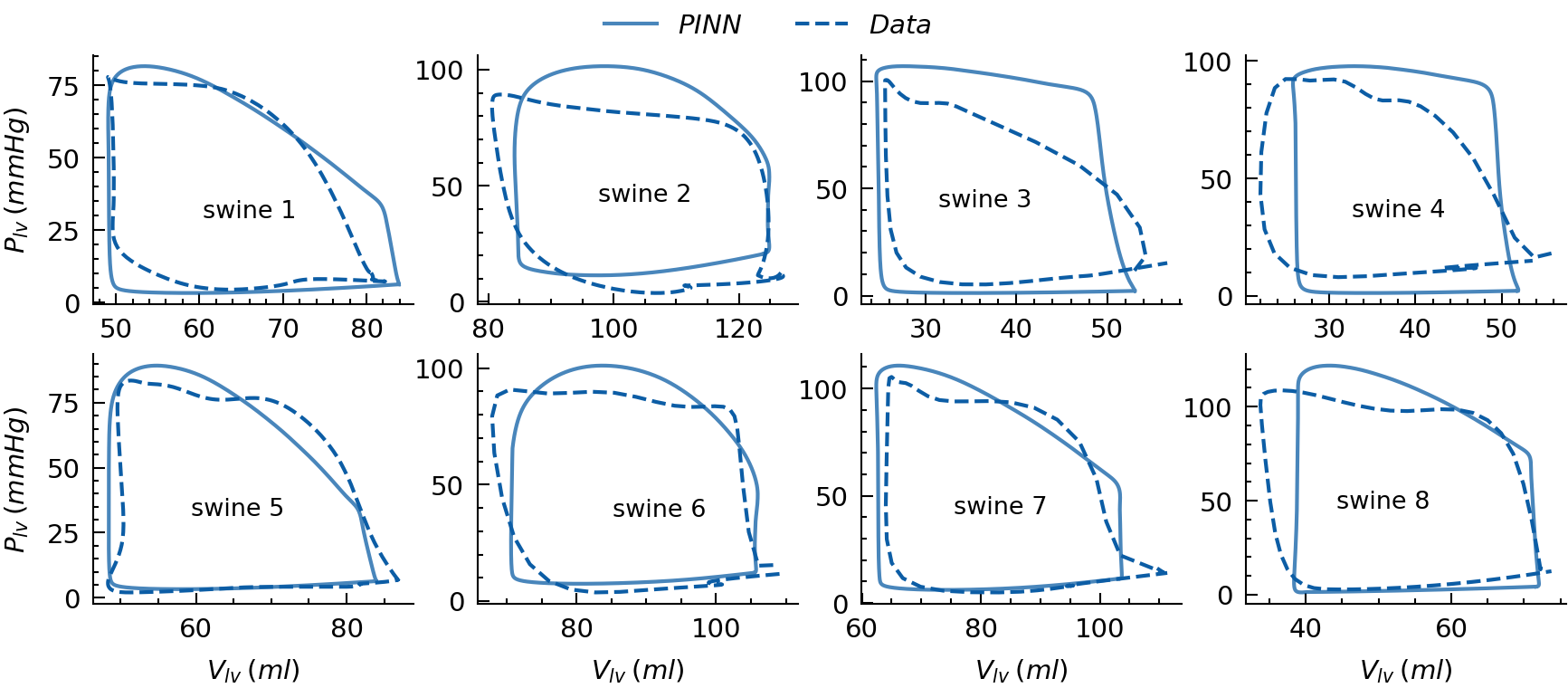}
      \caption{PV loops for eight swine subjects at their baseline}\label{fig:InvModelingBasePVloop}
    \end{subfigure}
  
    \begin{subfigure}[b]{0.85\textwidth}
      \includegraphics[width=\textwidth]{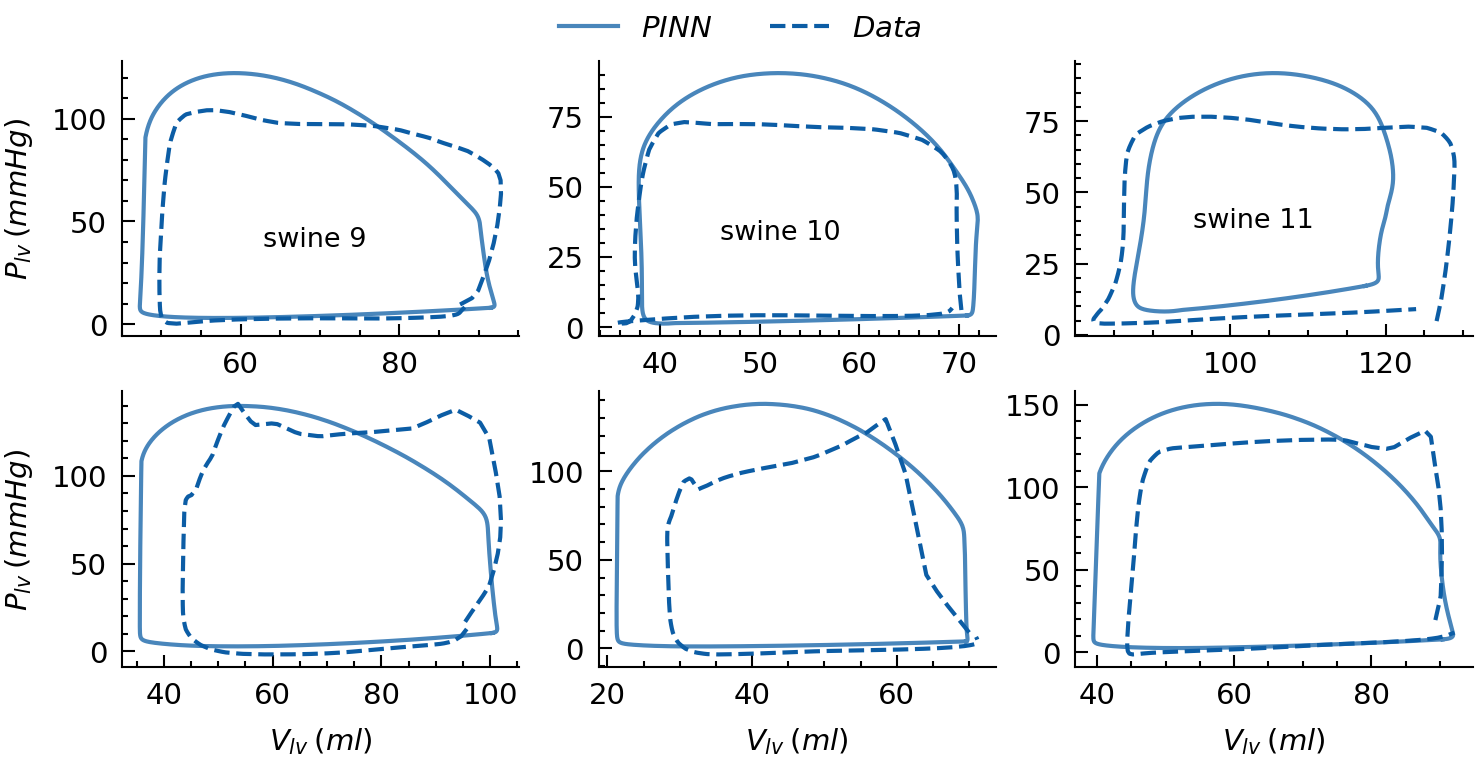}
      \caption{PV loops for three swine subjects before (top panel) and after (bottom) Dobutamine administration}\label{fig:InvModelingBasevsDbtmnPVloop}
    \end{subfigure}
    \caption{Comparison of the PV loops from inverse modeling with the corresponding measurements from the swine models}
    \label{fig:pvloop}
  \end{figure}

\end{appendices}

\end{document}